# Light effect on the interface resistance of $Ba_{0.8}Sr_{0.2}TiO_3$/$LaMnO_3$ heterosctructure

A.V. Leontyev[1], D.K. Zharkov[1], D.P. Pavlov[1], N.N. Garig'yanov[1], V.M. Mukhortov[2], D.A. Shulyaev[3], R.F. Mamin[1,4], V.V. Kabanov[1,5], V.V. Samartsev[1]

[1]*Zavoisky Physical-Technical Institute, FRC Kazan Scientific Center of RAS, 420029 Kazan, Russia*

[2]*Southern Scientific Center, Russian Academy of Sciences, 344006 Rostov-on-Don, Russia*

[3]*National University of Science and Technology "MISiS", 119991 Moscow, Russia*

[4]*Institute of Physics, Kazan Federal University, Kremlevskaya 16, 420008 Kazan, Russia*

[5]*Department for Complex Matter, Jozef Stefan Institute, 1000 Ljubljana, Slovenia*

**Abstract**

The effect of green and ultraviolet laser light on $Ba_{0.8}Sr_{0.2}TiO_3$/$LaMnO_3$ heterostructure electrical resistance was studied. In 80-200 K range illumination induces transient resistance component at ~15% of the steady-state value, the time constant associated with the transient component is ~12 s. The negative photoresistance effect is found under green, infrared and ultraviolet illumination.

**Introduction**

Unique properties of functional materials are achieved due to the effects associated with the complex composition of the interface structure. Such new materials include oxide heterointerfaces between two insulating oxides with unique transport properties due to strong electronic correlations between electrons. A high-mobility electron gas has been discovered at the interface between two oxide insulators $LaAlO_3$ (LAO) and $SrTiO_3$ (STO) by Ohtomo and Hwang in 2004 [1]. After that, this type of heterointerfaces involving two insulating nonmagnetic oxides were comprehensively studied. These investigations have attracted significant attention [1-12] due to a wide range of other physical phenomena observed in these systems. In particular, it was found that the metallic phase, so-called quasi-two-dimensional electron gas (q2DEG), is formed in the STO layers at the LAO/STO interface when the number of LAO layers is larger than three [3, 4]. Such a system undergoes a transition to a superconducting state below 300 mK [4]. Later on, the two-dimensional electron gas (2DEG) was revealed at interfaces between other nonmagnetic insulators, e.g., $KTaO_3/SrTiO_3$ [5] and $CaZrO_3/SrTiO_3$ [6]. The q2DEG was also found at the interfaces between magnetically ordered Mott insulators, in particular, at those involving ferromagnetic $GdTiO_3$ [7], as well as at the interface with antiferromagnetic $SmTiO_3$ [8] and $LaTiO_3$ [9], having the highest possible charge carrier density equal to $3 \times 10^{13}$ cm$^{-2}$. In addition, the LAO/STO heterostructure exhibits ferromagnetism [10]. It has been shown that analogously to the ionic polar discontinuity, the q2DEG [1] may be created at an interface due to electric polarization discontinuity [1, 2, 11, 12]. The creation of quasi-two-dimensional electron gas states at the interface and the ability to control such states by magnetic and electric fields is impossible without the use of new materials and without a development of a new design of the interfaces. We check the idea, that antiferromagnetic $LaMnO_3$ might be transferred to q2DEG state by increasing the concentration of free carriers at the interface with ferroelectric film [13]. This means that increasing the free charge carrier density may lead to a local ferromagnetic state in a system with q2DEG. Therefore, there is an opportunity to switch conductivity by an electric field in $Ba_{0.8}Sr_{0.2}TiO_3/LaMnO_3$ heterostructures.

The polar discontinuity at the interface, which leads to a divergence of the electrostatic potential, is usually discussed in order to describe the q2DEG formation [1, 12]. One of the most important features related to the q2DEG formation is the local polarity of layers inside the LAO slab. Here we choose the heterostructures like $BaTiO_3/LaMnO_3$ (BTO/LMO), where all layers are insulators and are neutral. On the other hand, these layers have a ferroelectric polarization due to Ti atoms displacements out of octahedron center in the BTO slab. The direction of the polarization may be switched by an external electric field. The numerical simulations of structural and electronic characteristics of the $BaTiO_3/LaMnO_3$ ferroelectric-antiferromagnet heterostructure had been performed previously [13]. Based on first-principles band structure calculations, we demonstrate the possibility of q2DEG formation at the interface composed of perovskite ferroelectric BTO and antiferromagnet manganite LMO [13, 14]. We present here new experimental results of the resistivity and photoresistivity measurements and of the effect of infrared, green and ultraviolet laser light illumination of the $Ba_{0.8}Sr_{0.2}TiO_3/LaMnO_3$ (BSTO/LMO) heterostructure.

The temperature dependence of the electrical resistance has been measured for heterostructures formed by antiferromagnetic $LaMnO_3$ single crystals and epitaxial films of ferroelectric $Ba_{0.8}Sr_{0.2}TiO_3$ (BSTO) deposited on the surface of single crystal in the form of parallelepiped of the size 4.0×3.2×0.7 $mm^3$ by reactive sputtering of stoichiometric targets using RF plasma (RF-sputtering) method [14, 15] at 650 C. The choice of these objects is primarily justified by the well-developed technologies used in the preparation of each individual component [15]. $Ba_{0.8}Sr_{0.2}TiO_3$ films have the ferroelectric transition temperature about 540 K, for 300-nm thick films on MgO substrate [15, 16]. We use 400-nm thick films because the properties of q2DEG are independent of thickness of the ferroelectric film above certain threshold. BSTO film on LMO substrate will be in ferroelectric state below room temperature. In this work, we present the photostimulated properties of the interface between ferroelectric oxide and insulating oxide in BSTO/LMO heterostructures.

**Experimental and results**

The interface resistivity was measured by the four-probe technique without and with illumination by infrared, green and ultraviolet laser light from the ferroelectric film side. The scheme of the experiment is presented in the inset in Fig. 1. Contact pads were created using silver paste connecting fine gauge golden wires to the surface of the sample. Two outer pads supplied current, other two inner pads were used to measure the voltage drop. The exact positions of the pads were different in different experiments. The sample was placed on a copper plate inside a Janis St-100 cryostat purged with dry nitrogen gas.

The temperature of the samples was measured with a copper-constantan thermocouple and was maintained with an accuracy of ± 0.1 K during the measurements. The temperature was stabilized using a Lakeshore 335 temperature controller. Light irradiation was carried out with an amplified Yb-doped fiber laser system producing pulses at 3 kHz (1.5 kHz and 0.75 kHz as well, as an option) with ~200-fs duration and mean power density of 80 mW·$(cm^{-2})$ at fundamental wavelength of 1028 nm (1.2 eV, infrared light), as well as doubled and quadrupled frequencies at 514 nm (2.4 eV, green light) and 257 nm (4.8 eV, ultraviolet light). The unfocused Gaussian laser beam was 4 mm in diameter, illuminating the space between the contact pads through an optical window of a cryostat.

The penetration depth $\lambda$ ($\lambda = 1/\alpha$, where $\alpha$ is an absorption coefficient) in BSTO (estimated from the absorption coefficient) is sufficiently greater than BSTO film thickness (400 nm) for IR (1.2 eV) and green (2.4 eV) light. Therefore, the substrate (LMO) states are also accessible for excitation. The bandgap of LMO is 0.5–1.3 eV, [21, 22], therefore infrared light illumination, creates free electron-hole pairs in LMO, affecting the measured resistivity. Ultraviolet light (4.8 eV) absorbs mostly within BSTO layer, but, in either case, photogenerated charge carriers are expected to contribute to the sample conductivity.

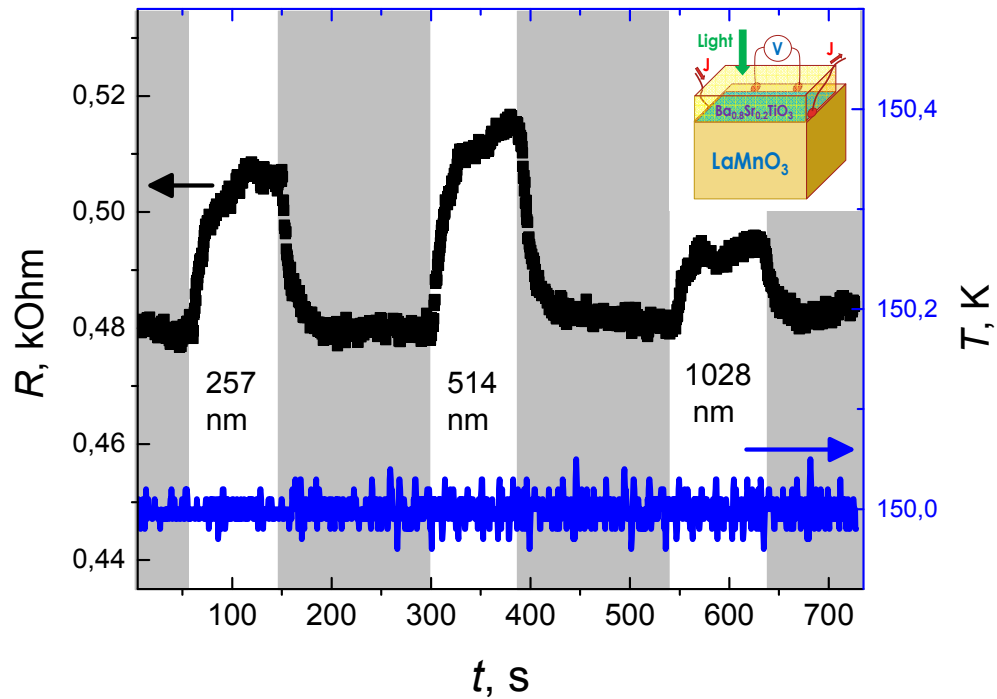

Fig. 1. Time dependency of the electrical resistance of $Ba_{0.8}Sr_{0.2}TiO_3/LaMnO_3$ heterostructure under infrared (the wavelength is 257 nm), green (514 nm) and ultraviolet (1028 nm) illumination (top curve). Dark periods are shaded. The sample temperature is stabilized at 150 K (bottom curve). Scheme of the experiment is presented in the insert.

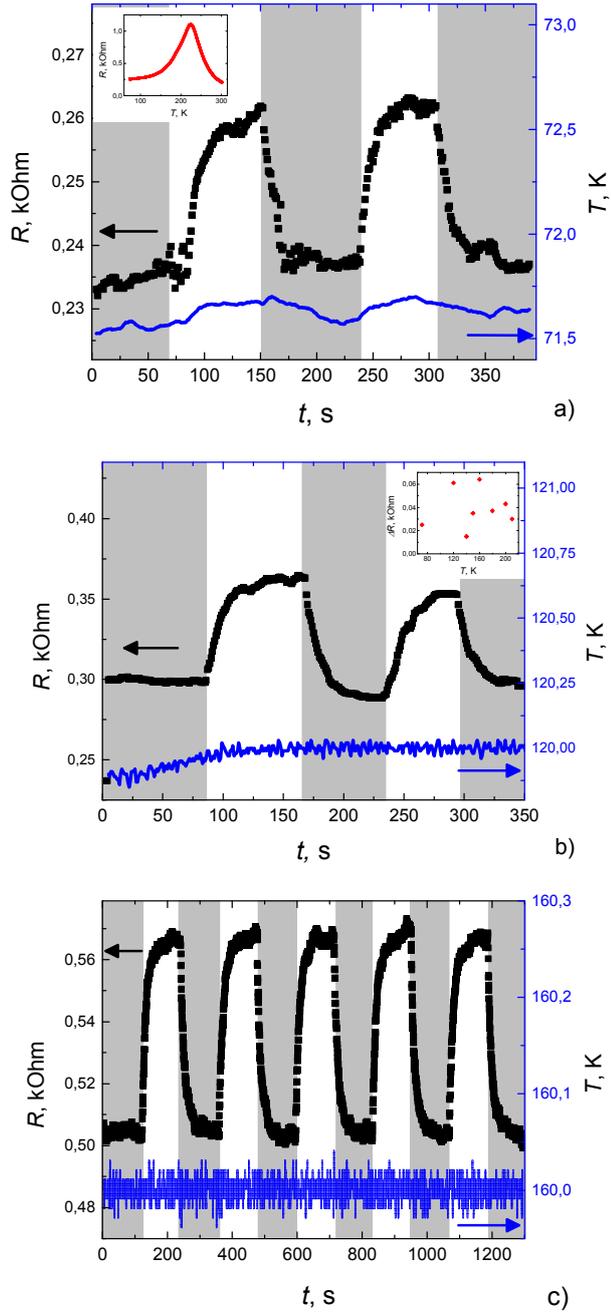

Fig. 2. Time dependency of the electrical resistance of $Ba_{0.8}Sr_{0.2}TiO_3/LaMnO_3$ heterostructure at 71.6 K (Fig. 2a), at 120.0 K (Fig. 2b) and at 160.0 K (Fig. 2c). Temperature dependence of the electrical resistivity of $Ba_{0.8}Sr_{0.2}TiO_3/LaMnO_3$ heterostructure is shown in the insert at Fig. 2a. The shift of the electrical resistivity of the $Ba_{0.8}Sr_{0.2}TiO_3/LaMnO_3$ heterostructure at different temperatures is shown in the insert at Fig. 2b.

We investigate the effect of illumination on $Ba_{0.8}Sr_{0.2}TiO_3/LaMnO_3$ heterostructure electrical resistance below 230 K, where the heterostructure demonstrate high conducting behavior. We conducted time-resolved photoresponse experiments by periodically turning the illuminating laser on and off and recording the response signal. In our case the response signal was the resistance change.

In Fig.1 the time evolution of photoresistance at 150 K under green, infrared and ultraviolet illumination is shown. An increase in resistance upon exposure to all of the used wavelengths, and the resistance recovery in dark state were observed. The response to green light illumination is the strongest. On the other hand, the response to IR light is considerably weaker.

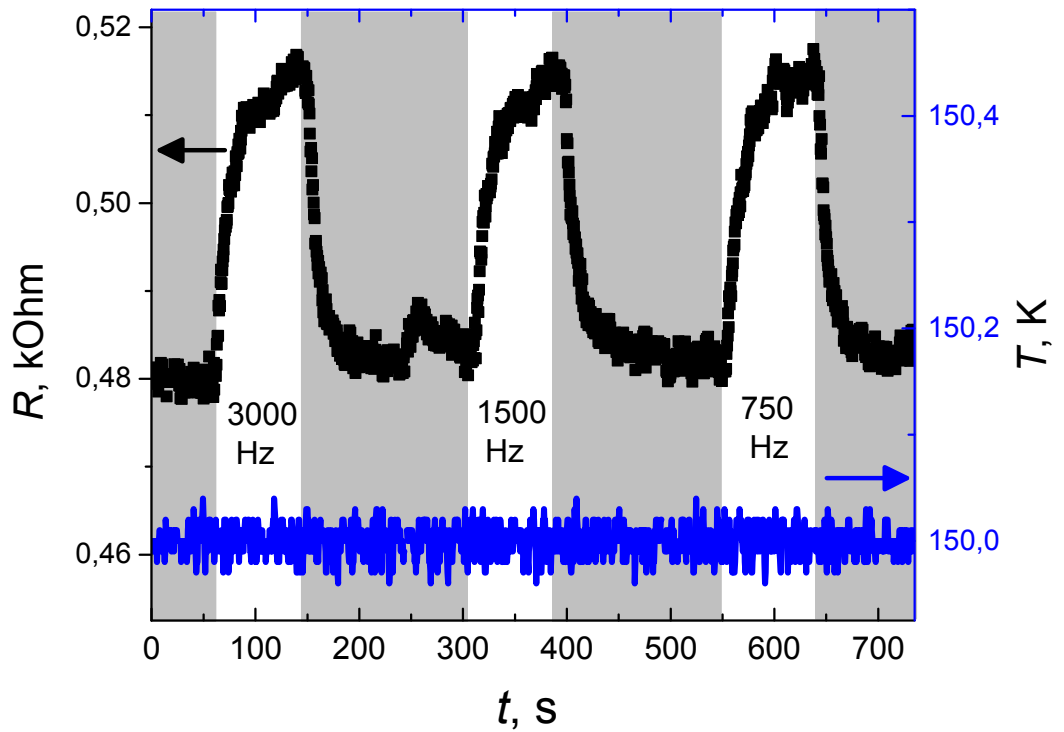

Fig. 3. Multiple on-off switchings of the electrical resistance of $Ba_{0.8}Sr_{0.2}TiO_3/LaMnO_3$ heterostructure at 150 K by green illumination (dark periods are shaded) with pulse rate at 3 kHz, 1.5 kHz and 0.75 kHz.

Temperature dependence of electrical resistivity of the $Ba_{0.8}Sr_{0.2}TiO_3/LaMnO_3$ heterostructure is shown in the insert at Fig. 2a. The time evolutions of photoresistance under green illumination at temperatures of 71.6 K, 120.0 K and 160.0 K are shown in Fig. 2 (Figs. 2a, 2b and 2c, correspondingly). The resistance rises under green illumination and recovers to the original values in the dark state for all temperatures. The effect of the green light on the electrical resistance of the Ba$_{0.8}$Sr$_{0.2}$TiO$_3$/LaMnO$_3$ heterostructure at different temperatures is shown in the insert in Fig. 2b. Such scattered values for different temperatures seem to indicate the influence of prehistory on the effect.

Multiple on-off switchings of electrical resistance of the Ba$_{0.8}$Sr$_{0.2}$TiO$_3$/LaMnO$_3$ heterostructure by green illumination at 150 K are shown in Fig. 3. Let's discuss whether the accumulative thermal effects of laser pulses can make a sizable contribution to the observed signal. Single ultrashort laser pulse has a power of the order of 100 MW cm$^{-2}$ and must lead to some instant sample heating. The overall temperature increase caused by the light have to be 5-10 K to explain the effect in this way. The thermalization time of photoexcited carriers is no longer than a few hundred femtoseconds and the recombination of trapped carriers occurs within a time scale of tens of picoseconds [20]. Any heat transferred to the lattice is being dissipated into the substrate, which is mounted on the cryostated copper plate, within tens of nanoseconds. On the other hand, the interval between consequent laser pulses is few hundreds of microseconds. Nevertheless, to test this possibility experimentally we have conducted the same experiments with two-fold and four-fold time intervals between pulses (so, 1.5 kHz and 0.75 kHz in addition to 3 kHz, Fig. 3, and the intensity of laser pulses was adjusted in order to keep average illumination dose constant). If the sample does not cool down fully before the next pulse, it would add up to the temperature with every pulse, and changing the pulse interval has to have major effect on heating rate. However, Fig. 3 shows no effect of laser pulse repetition rate on photoresistance, neither its maximum value, nor increase/decrease rate.

Let us also make an estimate from general considerations how much the sample can heat up. Laser heating of the sample is an important experimental problem in these experiments and has often been discussed in the literature [25-29]. The steady-state temperature of the near surface volume increases due to the pump laser excitation. The thickness of the substrate (~ 0.7 mm) and even of ferroelectric film (~ 350 nm) is typically much larger than the absorption length (~ 80 nm). Thus we can calculate the temperature rise using a simple steady-state heat diffusion model [25, 27], where the Gaussian laser beam with the average laser power is focused into a spot of diameter $d$. In our experiment we use unfocused laser beam ($d$~ 4 mm). Therefore, the change of the temperature on the surface $\Delta T$ is about 0.1 K, according our estimations using data for absorption coefficient and thermal conductivities of LMO [30-32]. Our estimation correlate with other estimations when for focused laser beam ($d\sim 0.1$ mm) $\Delta T$ is about 12 K or 2.7 K for much higher mean power density [28, 29, 33]. Thus heating effect at the interface area is nearly absent in our case ($\Delta T$ is ~ 0.1 K on the surface) because we use unfocused laser beam and the mean power density was relatively very small.

The resistance of the heterostructure decreases with temperature decreasing demonstrating a high conducting behavior below the temperature about 230 K [17] (see the inset in Fig. 2a). The emergence of a region of high conductivity cannot be explained, for example, by the positive temperature coefficient of the effect of resistivity of the ferroelectric film [18, 19]. This effect can change a temperature dependence of BSTO film resistivity due to depletion layers formed at the interfaces of ferroelectrics. But the resistance of the BSTO film is so huge, that it is shunted by the LMO substrate, because the current flow area is much larger in the substrate, and also the LMO resistivity (see the inset in Fig. 1) is smaller than the BSTO resistivity. Therefore, we cannot observe the resistance or any change in the resistance of the BSTO film in our experiments. Thus, we believe the transition to the state with high conductivity at the interface occurs [13, 17].

Finally, we propose the following explanations of the observed phenomena. Since we observe rather slow relaxation processes, we assume that these charge dynamic is associated with the relaxation in the ferroelectric film (relaxation processes in the metallic part of the sample should be much faster). Under illumination, the carriers generated in the ferroelectric will screen the ferroelectric polarization. Therefore, this reduction of bulk polarization in the ferroelectric film reduces the carrier density in the interface region. To clarify all these issues, further experiments with switching the polarization of the ferroelectric film in strong electric fields are necessary.

**Conclusion**

At low temperatures (T<~210 K) as the illumination starts, the electrical resistance begins to grow with a time constant of ~12 s and then it reaches the steady state value while the sample is continuously illuminated. This behavior is opposite to usual photoconductivity in ordinary semiconductors where due to photo-carriers the conductivity increases. Thus the effect of the pulsed illumination on the $Ba_{0.8}Sr_{0.2}TiO_3/LaMnO_3$ heterostructure electrical resistance is observed. This effect allows to control interface conductivity by the laser light illumination. The negative photoresistance effect is detected for green, infrared and ultraviolet illumination. The observed effect cannot be explained by the direct heating of the sample by laser pulses, because the laser pulse repetition rate is low and therefore the cumulative thermal effects should be negligible. This example shows that quasi-two-dimensional high conductance at ferroelectric/dielectric interfaces can be controlled by relatively simple technique.

**Acknowledgements**

The reported study was funded by Russian Science Foundation according to the research project No. 18-12-00260. Numerical calculations were implemented with financial support of RFBR (project No. 20-02-00545). R. F. M. acknowledges partial support by the Russian Government Program of Competitive Growth of Kazan Federal University. V. V. K. acknowledges financial support from Slovenian Research Agency Program P1-0040.

Author's address:

Dr. R.F.Mamin,

Kazan Physical-Technical Institute of

Russian Academy of Sciences,

Sibirskii trakt 10/7

420029 Kazan, Russia

E-mail: mamin@kfti.knc.ru.

Fax: (843) 272-50-75